# Ultrafast, Zero-Bias, Graphene Photodetectors with Polymeric Gate Dielectric on Passive Photonic Waveguides

Vaidotas Mišeikis,◇ Simone Marconi,◇ Marco A. Giambra, Alberto Montanaro, Leonardo Martini, Filippo Fabbri, Sergio Pezzini, Giulia Piccinini, Stiven Forti, Bernat Terrés, Ilya Goykhman, Louiza Hamidouche, Pierre Legagneux, Vito Sorianello, Andrea C. Ferrari, Frank H. L. Koppens, Marco Romagnoli,* and Camilla Coletti*

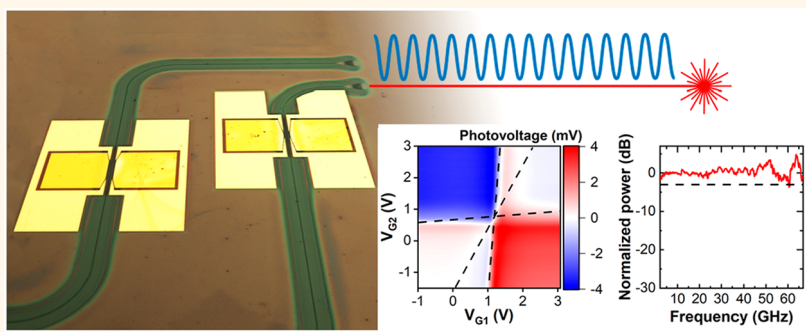

**ABSTRACT:** We report compact, scalable, high-performance, waveguide integrated graphene-based photodetectors (GPDs) for telecom and datacom applications, not affected by dark current. To exploit the photothermoelectric (PTE) effect, our devices rely on a graphene/polymer/graphene stack with static top split gates. The polymeric dielectric, poly(vinyl alcohol) (PVA), allows us to preserve graphene quality and to generate a controllable p−n junction. Both graphene layers are fabricated using aligned single-crystal graphene arrays grown by chemical vapor deposition. The use of PVA yields a low charge inhomogeneity $\sim 8 \times 10^{10}$ cm$^{-2}$ at the charge neutrality point, and a large Seebeck coefficient $\sim 140$ $\mu$V K$^{-1}$, enhancing the PTE effect. Our devices are the fastest GPDs operating with zero dark current, showing a flat frequency response up to 67 GHz without roll-off. This performance is achieved on a passive, low-cost, photonic platform, and does not rely on nanoscale plasmonic structures. This, combined with scalability and ease of integration, makes our GPDs a promising building block for next-generation optical communication devices.

Telecommunication networks and interconnections in data centers require a bandwidth increase combined with a reduction of power consumption and cost to cope with the growing demands for data transmission.[1,2] The Ethernet roadmap[3] foresees a bandwidth doubling roughly every two years. The present target is to develop transceiver (transmitters and receivers) modules working at 1.6 Tb s$^{-1}$ by 2022.[4] The established technologies based on InP[5] and Si photonics[6] are continuously improving. However, the requirements in terms of bandwidth and power consumption have not been fulfilled in one system yet.[6]

Single-layer graphene (SLG) is ideally suited for optoelectronic and photonic applications.[7−10] The absence of a bandgap enables absorption in a very broad range of optical frequencies, spanning from the UV to the far-infrared.[7,9,11] The ultrafast electron excitation dynamics (< 50 fs)[12] upon optical excitation, and the consequent short relaxation time, of the order of few ps,[12] enable high-speed devices.[13] Fast graphene photodetectors (GPDs) have been demonstrated,[8,14−19] with bandwidth > 100 GHz.[17−19] In terms of speed, such devices can compete with Ge-based PDs.[20] However, these GPDs typically require nanoscale plasmonic structures, using metals not compatible with complementary metal oxide semiconductor (CMOS) integration.[18,19] Furthermore, these GPDs are based on photovoltaic (PV)[19] and photobolometric (PBM)[17,18] effects and are thus operated with a bias ∼0.5 V[17,18] to ∼1.5 V,[19] leading to significant dark current (up to a few mA),[18] orders of magnitude larger than typical p−n junction near-infrared PDs.[21]

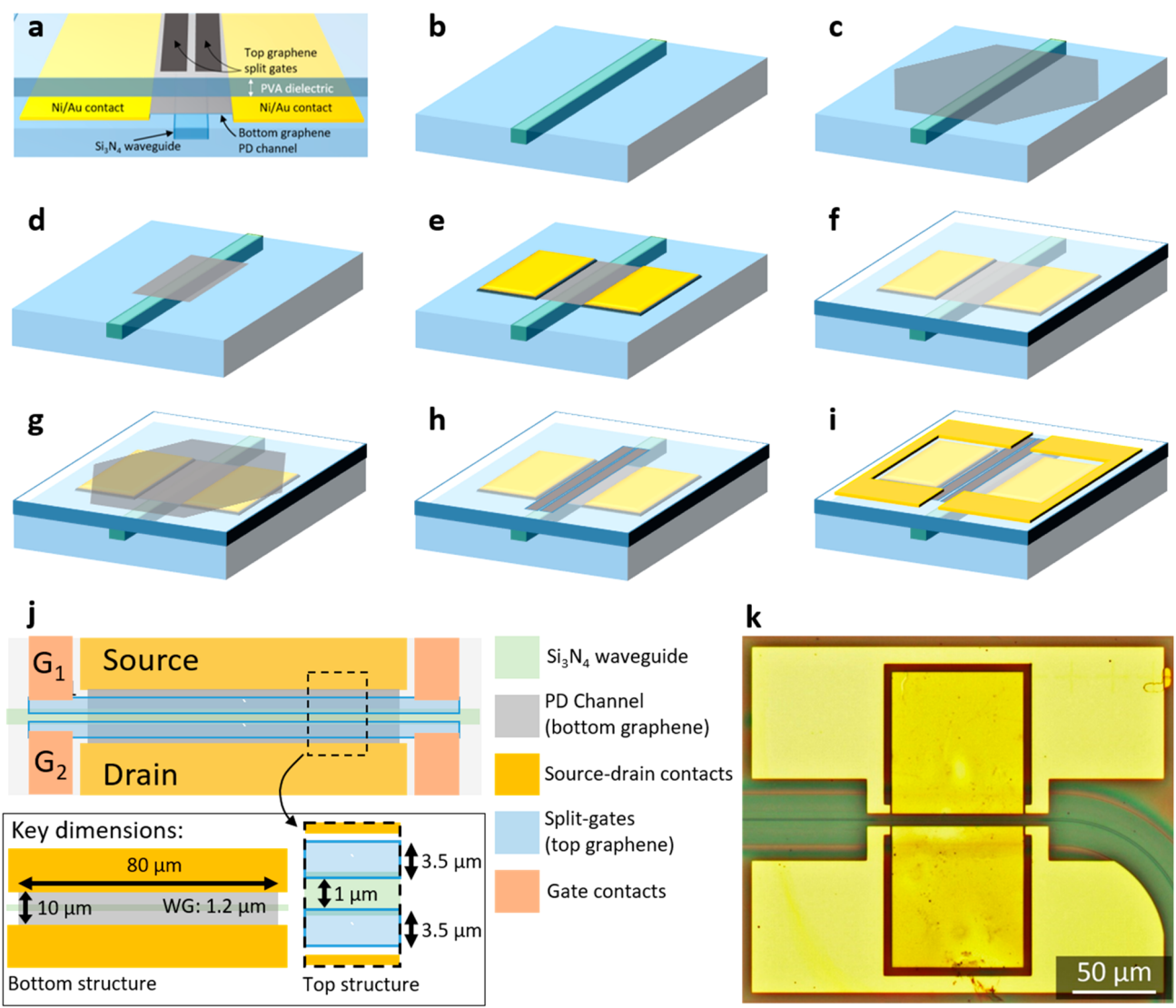

Figure 1. Design and fabrication flow of double SLG PTE PD. (a) Cross-section. (b) Wafer (light blue) with a $Si_3N_4$ photonic WG (teal). (c) Single-crystal SLG (gray) transferred on the WG. (d) SLG shaped into PD channel. (e) Ni/Au contacts (yellow) deposited by thermal evaporation. (f) PVA dielectric (semi-transparent blue) spin-coated on the chip. (g) Top SLG (darker gray) aligned and transferred on the device. (h) Top SLG shaped into split gate geometry by RIE. (i) Ni/Au gate contacts (darker yellow) deposited by thermal evaporation. (j) Schematic diagram of design (not to scale) dimensions optimized for high $R_V$. Inset: key dimensions of bottom and top SLG structures. (k) Optical micrograph of a typical device. Darker yellow squares are the contact pads for the GPD channel (not visible). Lighter yellow are the pads for the split gates. A $Si_3N_4$ WG (thin dark green line) is visible at the center of the device.

GPDs based on the photothermoelectric (PTE) effect promise large bandwidth,[8,22] with the additional advantage of bias-free operation and, thus, zero dark current photovoltage generation.[15,23−25] The absence of dark current avoids any noise contribution coming from a DC current.[26] Indeed, for a large (∼mA) dark current, we expect an increase of generation−recombination (GR) noise caused by the statistical generation and recombination of charge carriers.[27] The GR noise power spectral density (PSD) is directly proportional to the square of the DC current.[28] GR noise contributes to the overall noise also at microwave frequencies (2.5 GHz).[29] With a dark current of the order of mA, even shot noise, originating from the discrete nature of the electric charge, and dependent on the current flowing without relation to operating temperature[27] and with PSD proportional to the DC current,[29] may give a non-negligible contribution.[29] In PTE-based GPDs, the photoresponse is generated by the Seebeck effect induced by two main factors: the spatial gradient of the SLG electronic temperature induced by the absorption of the optical signal, and the spatial variation of the SLG Seebeck coefficient, $S$.[30] The $S$ profile along SLG can be induced by electrostatically generated p−n junctions.[31] Such junctions can be achieved either by doped waveguides (WGs),[15,25] with increased fabrication complexity and cost, or by external gates,[23] which requires SLG dielectric encapsulation. The photovoltage $V_{ph}$ can be written as[30−32]

$$V_{ph} = (S_2 - S_1)(T_e - T_0) = \Delta S \Delta T_e \quad (1)$$

where $S_2$ and $S_1$ are the Seebeck coefficients in each side of the SLG p−n junction, $T_e$ is the electron temperature, and $T_0$ is the SLG lattice temperature. The ratio between $V_{ph}$ and the incident optical power $P_{opt}$ is the voltage external responsivity, $R_V$, expressed in V W$^{-1}$.[15,24,25]

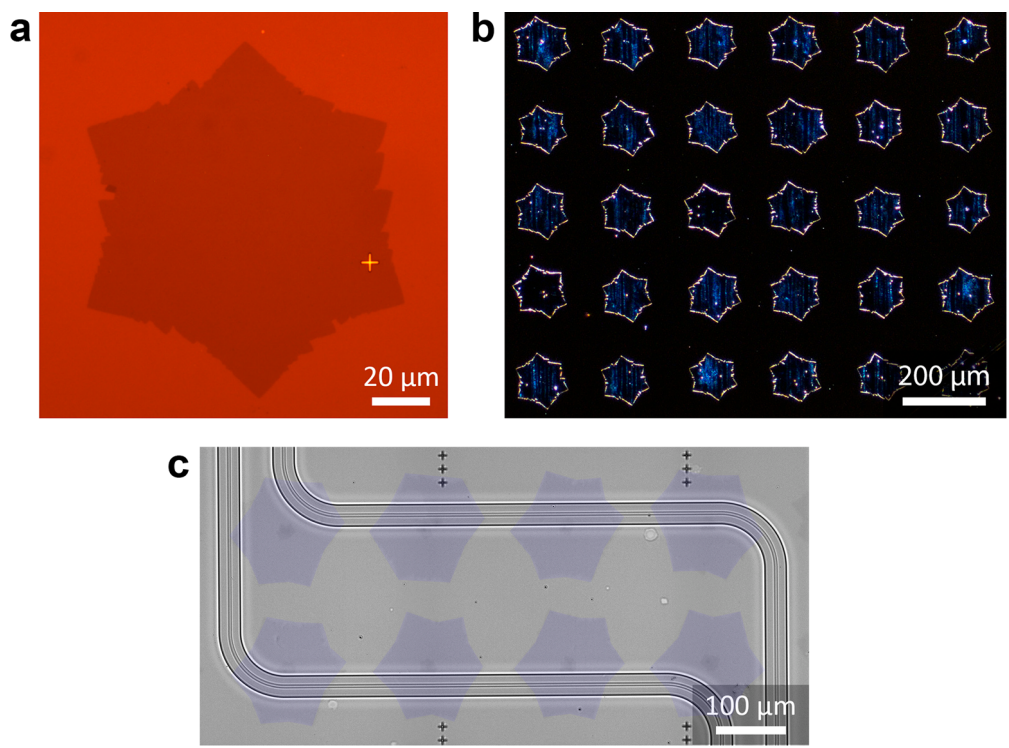

Figure 2. (a) Optical image of a typical SLG crystal transferred on 285 nm SiO$_2$. (b) Dark-field microscopy image of a SLG crystal array with a periodicity of 200 $\mu$m. (c) False color optical image of SLG array transferred on two adjacent WGs.

Waveguide-integrated PTE-GPDs were reported,[15,23−25] with $R_V \sim 3.5$ V W$^{-1}$ (ref 25) to 12.2 V W$^{-1}$ (ref 24) but limited in bandwidth (BW), with evidence of roll-off at 65 GHz for exfoliated SLG,[25] and at 42 GHz for CVD SLG.[24] These performances were reached by enhancing the optical absorption, exploiting subwavelength confinement of the electromagnetic field with nanoscale structures, such as photonic crystals,[15] slot,[25] and plasmonic WGs.[24] The local field enhancement in SLG allows for a higher gradient of the electronic temperature $\Delta T_e$, leading to a larger photovoltage, but the fabrication of these structures typically requires sub-100 nm resolution, therefore a simpler GPD geometry based on a passive straight WG is preferable.

According to eq 1, the photovoltage may also be improved by increasing $S$. This is related to the SLG mobility, $\mu$,[24] and residual carrier concentration $n^*$ at the charge neutrality point (CNP).[33] To date, a large $S$ (i.e., $S \sim 183$ $\mu$V K$^{-1}$) was only reported for exfoliated SLG on hexagonal boron nitride (hBN) ($\mu > 10,000$ cm$^2$ V$^{-1}$ s$^{-1}$).[34] For CVD grown polycrystalline SLG, $S < 20$ $\mu$V K$^{-1}$ was observed.[35]

Here, we demonstrate a SLG-polymer-SLG PTE GPD fabricated using scalable CVD single-crystal SLG arrays for both channel and split gates. The use of polymer dielectric, poly(vinyl alcohol) (PVA) allows us to preserve the SLG quality (i.e., $n^* = 8 \times 10^{10}$ cm$^{-2}$ at the CNP and $\mu \sim 16,000$ cm$^2$ V$^{-1}$ s$^{-1}$) and to obtain $S$ up to 140 $\mu$V K$^{-1}$. The GPDs have a compact footprint and are fabricated on a single-mode straight WG using top gates to electrostatically induce a p−n junction, thus not requiring the ion implantation used in conventional devices.[26] Deposition of common dielectrics, like Al$_2$O$_3$, HfO$_2$, and Si$_3$N$_4$, typically leads to a degradation of $\mu$ and $n^*$ (refs 36−40). Here, PVA is deposited by spin-coating, yielding a conformal coverage. It is nonsoluble in organic solvents,[41] widely used in nanofabrication, and does not compromise the SLG properties. Static characterization at 1550 nm shows a 6-fold pattern photovoltage map as a function of gate voltage, a signature of PTE.[32] The responsivity is ∼6 V W$^{-1}$, comparable to GPDs using photonic structures with feature sizes <100 nm.[15,24,25] Dynamic characterization is performed up to 67 GHz, showing a flat electro-optical frequency response without roll-off. This frequency response is, to the best of our knowledge, the highest thus far for a zero-bias GPDs.

## RESULTS

A schematic cross-section diagram of GPDs is in Figure 1a. They comprise a CVD single-crystal SLG channel on a Si$_3$N$_4$ WG, a PVA dielectric spacer, and top SLG split gates above the SLG channel, aligned with the center of the photonic WG. By applying a voltage to each split gate, a p−n junction is created enabling the generation of photovoltage when a $T_e$ gradient is induced across the junction by light absorption in a SLG channel above the photonic bus WG.[30]

Figure 1b−i outlines the process flow. Devices are fabricated on a wafer containing Si$_3$N$_4$ WGs cladded (as depicted in panel b) with a thin (∼25 nm) layer of boron−phosphorus tetraethyl orthosilicate (BPTEOS) (further information in Methods). SLG crystals are deterministically grown in arrays,[42] matching

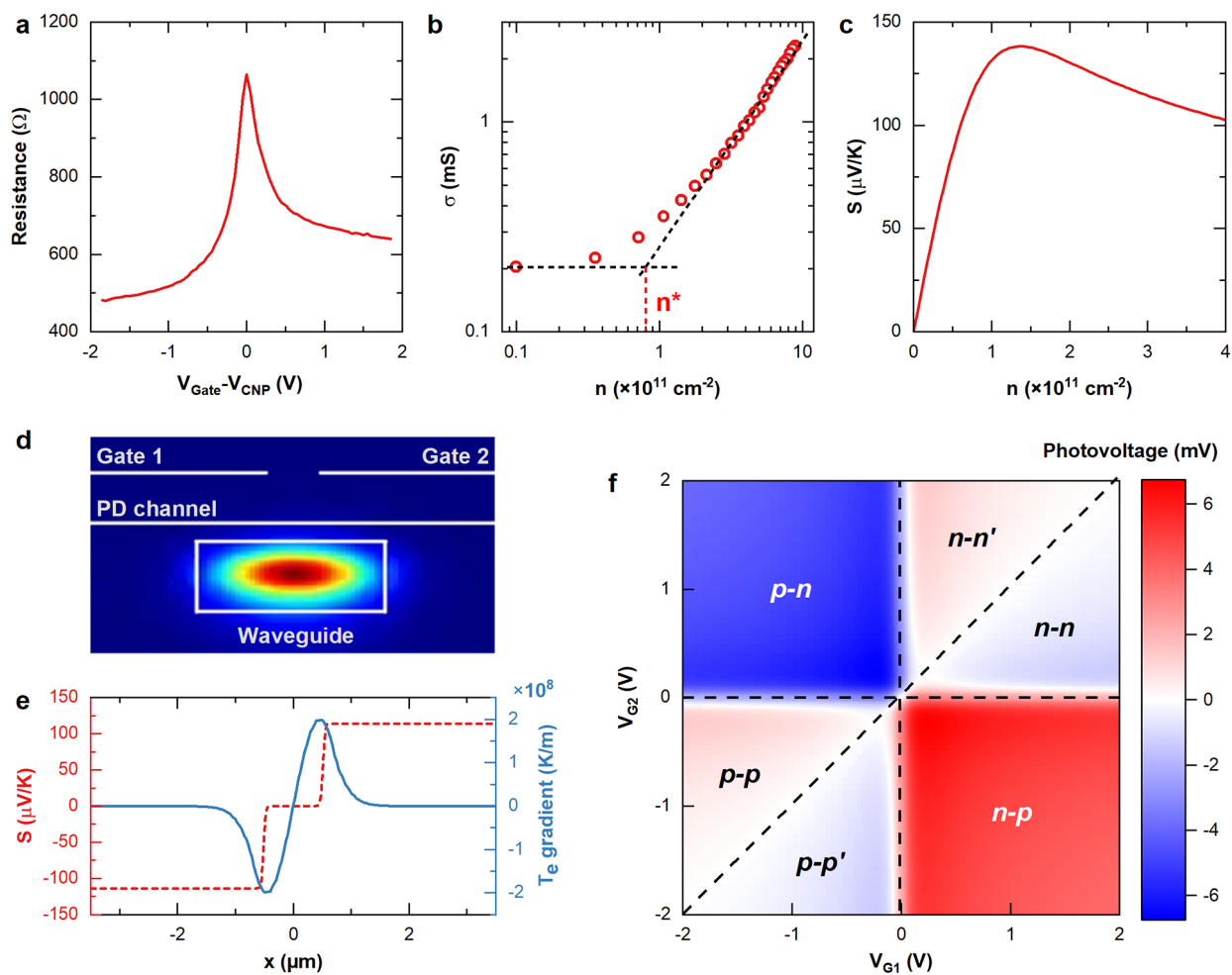

Figure 3. (a) Field-effect response of SLG/PVA. (b) Extraction of $n^*$. (c) Calculated $S$ for $n^* = 8 \times 10^{10}$ cm$^{-2}$. (d) Electric field profile of the simulated fundamental quasi-TE mode of the Si$_3$N$_4$ WG (1200 nm × 260 nm, $\lambda_0 = 1550$ nm). (e) $S$ spatial profile in SLG p−n junction (red curve) and spatial profile of $\Delta T_e$ (blue curve). (f) Simulated photovoltage map as a function of voltage applied to the split gates, assuming 1 mW incident power, for channel length 10 μm and channel width 80 μm (see Figure 1j). Different regions are evident, corresponding to different doping configurations, p−n, p−p, n−p, and n−n. As emphasized by the dashed lines, the photovoltage sign reversal leads to additional p−p′ and n−n′ configurations, thus the 6-fold symmetry characteristic of PTE.[30,32]

the WG geometry, electrochemically delaminated from a Cu via semidry transfer,[42] and aligned (panel c) on the WGs. Deterministic growth and transfer allows us to transfer up to 300 crystals at a time on adjacent WGs and to process ∼20 at a time. Images of the SLG crystals used for GPD fabrication are shown in Figure 2a−c. SLG is then shaped into the active device channels using electron-beam lithography (EBL) and reactive-ion etching (RIE), Figure 1d. Ohmic contacts are fabricated using EBL and thermal evaporation of 7 nm Ni and 60 nm Au, followed by lift-off in acetone, Figure 1e. A PVA solution (5% in water, Sigma-Aldrich, MW: 23−75 kDa, 89% hydrolyzed) is spin-coated on the devices at 8000 rpm, achieving a conformal coating. This is then cured on a hot plate at 90 °C for 2 min, Figure 1f. A second layer of SLG single-crystals is then aligned and laminated on the devices using semidry transfer,[42] Figure 1g. The top SLG crystals are shaped into split gates and contacted using the same methods used for the bottom SLG, Figure 1h,i. To ensure the operation of the gates even in the event of a discontinuity in the 80 μm-long and 3.5 μm-wide SLG stripes, metal connections to the gates are fabricated from both sides in a "butterfly" configuration, Figure 1i,k. As a final step, PVA is removed from the WGs outside the device with deionized (DI) water. The deterministically grown single-crystal SLG arrays and the PVA dielectric allow us to make multiple devices in parallel. PVA, if left uncovered, could be affected by humidity.[43] A further encapsulation step via semidry transfer of CVD hBN can be used to provide a humidity barrier to the PVA.

In order to evaluate both the SLG electrical and thermoelectric properties when PVA is used as the gate dielectric, $n^*$ is extracted from the field-effect transistor (FET) measurements (see Methods). A typical field-effect curve for a SLG FET using PVA as gate dielectric is in Figure 3a. We get $n^* \sim 8.2 \times 10^{10}$ cm$^{-2}$ (∼35 meV) by performing a linear fit (Figure 3b) to a logarithmic plot of conductivity as a function of carrier density.[44,45] A similar value is also obtained by fitting the field-effect resistance curves as for ref 46 (see Methods for details). This $n^*$ is lower than in samples encapsulated using technologies compatible with wafer-scale processing, such as atomic layer deposition (ALD) (∼2.3 × 10$^{11}$ cm$^{-2}$ [46] and ∼3 × 10$^{11}$ cm$^{-2}$[38]) and is approaching those reported for exfoliated or CVD-based SLG/hBN heterostructures.[44,45,47,48]

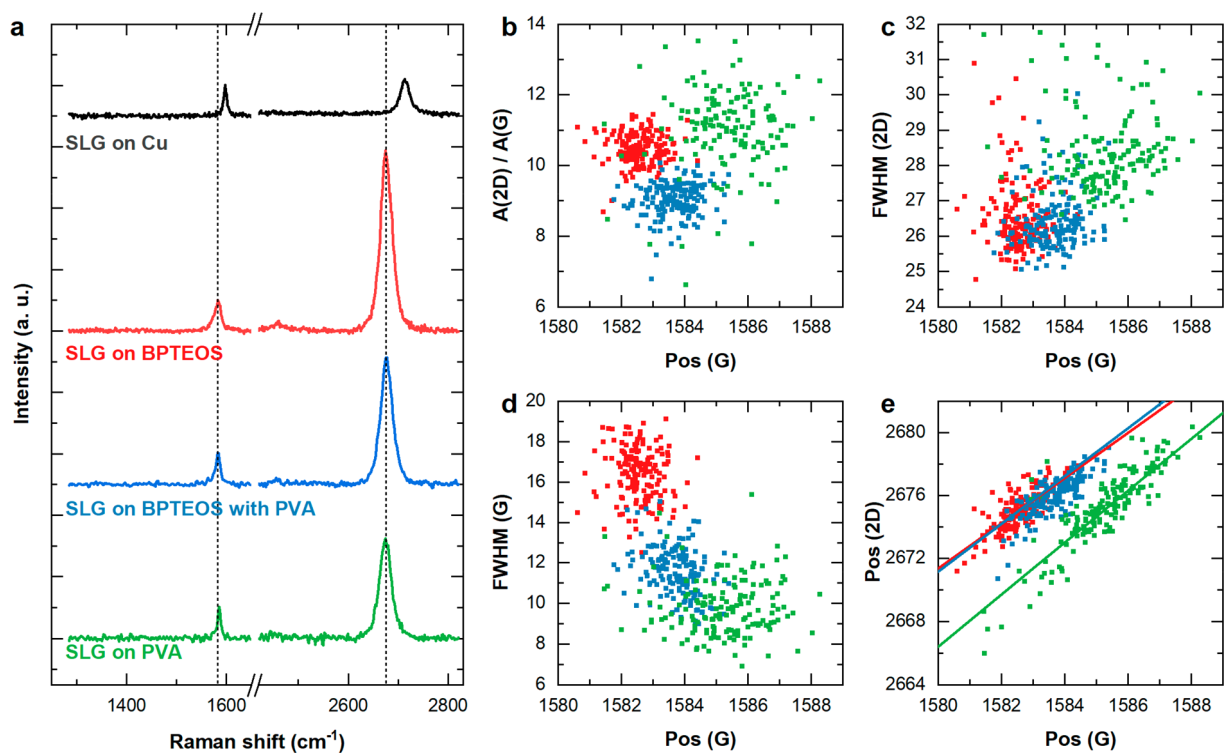

**Figure 4.** (a) Representative spectra for SLG grown on Cu (gray), transferred on BPTEOS (red), with PVA coating (blue) and on top of PVA (green). The spectra are normalized to have the same I(G). Same colors are used in the correlation plots (b−d). (b) $A(2D)/A(G)$ as a function of Pos(G). (c) FWHM(2D) as a function of Pos(G). (d) FWHM(G) as a function of Pos(G). (e) Pos(2D) as a function of Pos(G). Solid lines of corresponding color show linear fits of the data for the three sample configurations.

We use the model of ref 33 to calculate $S$ using our experimental $n^*$. $S$ as a function of $n$ is plotted in Figure 3c (see Methods for details). We get $S \sim 140$ $\mu$V K$^{-1}$ for $n^* \sim 8 \times 10^{10}$ cm$^{-2}$. This is higher than for SLG on SiO$_2$[35,49−51] and similar to that reported for SLG on hBN.[34]

In order to model the PTE effect and to compute the $T_e$ profile along with the generated photovoltage, we adapt the model of ref 32 to describe a GPD on a photonic integrated WG. Figure 3d shows the electric field profile of the simulated fundamental quasi-TE mode propagating in a Si$_3$N$_4$ WG, using a commercial mode solver, and the surface conductivity model for the SLG optical properties of ref 52 (see Methods for further details). An example of the spatial profile of $\Delta T_e$ in a cut of the GPD is in Figure 3e, along with the $S$ spatial profile for two gate voltages.

The optimum design is chosen by simulating photovoltage maps of several device geometries with channel length, $L$, ranging between 10 and 25 $\mu$m and width, $W$, between 40 and 80 $\mu$m. The optimization of the aspect ratio, $W/L$, of the SLG channel is necessary to minimize the series resistance in order to reduce the electrical power dissipation of the GPD when connected to 50 $\Omega$ matched read-out electronics. The maximum $W$ is determined by the length of the mode absorption. Above 50 $\mu$m, the mode is attenuated by a factor $e$. Therefore, for devices with $W \gg 50$ $\mu$m, no significant increase in photocurrent is expected. Accordingly, the photovoltage, $V_{Ph}$, drops due to a saturated photocurrent, $I_{Ph}$, and decreasing resistance, $R$ (as per $V_{Ph} = R \times I_{Ph}$).[30] For a minimum channel length $L \sim 10$ $\mu$m, determined by fabrication constraints (as discussed in Methods), a channel width $W \sim 80$ $\mu$m provides the best compromise, with a resistance of several hundred $\Omega$ and $R_V > 5$ V W$^{-1}$. The gate electrodes, Figure 1j, consist of two SLG strips 3.5 $\mu$m wide separated by a 1 $\mu$m gap. This spacing is chosen as it is compatible with typical i-line optical lithography at 365 nm,[64] making it suitable for large-volume production. Even though EBL can achieve a resolution $\sim$50 times smaller than that of i-line optical lithography, simulations of devices having split gate separation in the range $\sim$0.2−1.1 $\mu$m show only a small penalty (22%) of maximum achievable $R_V$ when going from EBL to optical resolution (see Figure 11e). The photovoltage is simulated as a function of the split gate voltage, as shown in Figure 3f. The maximum photovoltage is observed in the regions with opposing gate polarity, n−p and p−n. Doping configurations where both gates are of the same polarity with respect to the CNP are split into additional regions n−n, n−n′ and p−p, p−p′, with n′ (p′) indicating stronger n-type (p-type) doping with respect to n (p). A photovoltage sign reversal is observed due to the nonmonotonic variation of $S$ with gate voltage.[30,32] This leads to a 6-fold pattern of alternating positive and negative photovoltage, typical of PTE.[30,32] Dashed guides to the eye separate the six distinct regions.

At each fabrication step, the SLG quality is assessed by Raman spectroscopy. Raman spectra are acquired at 532 nm with a Renishaw InVia spectrometer, a laser power $\sim$1 mW, and 50× objective, giving a spot size $\sim$2 $\mu$m. The top SLG layers Raman spectra are measured outside the areas with two overlapping SLG, allowing for an independent analysis of top gate and active channel.

Representative spectra of SLG crystals are shown in Figure 4a: SLG on Cu, SLG after transfer on BPTEOS, SLG after PVA coating, and SLG after transfer on PVA. The SLG spectrum on Cu (shown after Cu luminescence subtraction) has a 2D peak with a single Lorentzian shape and with a FWHM(2D) $\sim$ 22.4 cm$^{-1}$, a signature of SLG.[53] The G peak position, Pos(G), is

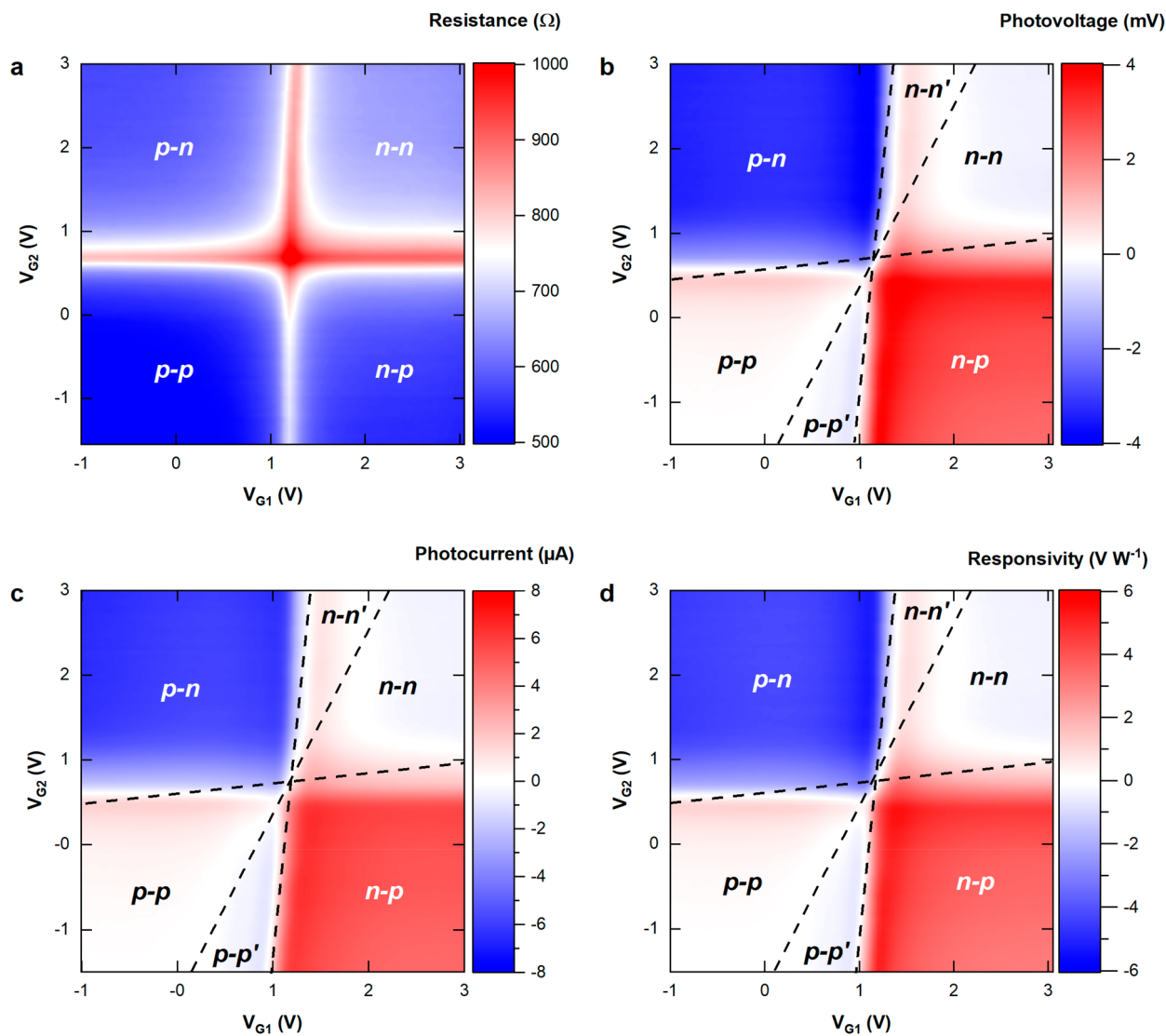

Figure 5. (a) Resistance map as a function of the voltage applied to split gates. Four regions are evident, corresponding to p−n, p−p, n−p, and n−n doping. (b) Photovoltage map. Sign reversal leads to the appearance additional regions p−p′ and n−n′, thus a six-fold symmetry, characteristic of PTE.[30,32] (c) Photocurrent and (d) responsivity maps as a function of voltage applied to both gates, showing six-fold symmetry.

~1597 cm$^{-1}$, with FWHM(G) ~ 8 cm$^{-1}$. The 2D peak position, Pos(2D) is ~2713 cm$^{-1}$, while the 2D to G peak intensity and area ratios, $I(2D)/I(G)$ and $A(2D)/A(G)$, are ~1.2 and ~1.3, respectively. No D peak is observed, indicating negligible defects.[54,55] After transfer on BPTEOS, the 2D peak retains its single-Lorentzian line shape with FWHM(2D) ~ 26.5 cm$^{-1}$. After PVA coating, FWHM(2D) ~ 26.4 cm$^{-1}$. The D peak remains negligible, indicating that no significant defects are induced by SLG transfer or PVA.

Raman mapping is performed over an area ~20 μm × 8 μm on SLG after transfer on BPTEOS, after PVA coating, and after transfer on PVA. Figure 4b−e plots Raman data extracted from the maps: $A(2D)/A(G)$, FWHM(2D), FWHM(G), and Pos(2D) as a function of Pos(G). Pos(G) depends on both doping[56−58] and strain.[59] This implies that local variations in strain and doping manifest as a spread in Pos(G) which, after transfer and PVA deposition, is ~1582.5 ± 0.6 cm$^{-1}$ and 1583.7 ± 0.7 cm$^{-1}$, respectively. For graphene on BPTEOS, FWHM(G) is ~16.4 ± 1.3 cm$^{-1}$, Pos(2D) ~2675.0 ± 1.2 cm$^{-1}$, $I(2D)/I(G)$ ~6.5 ± 0.5, and $A(2D)/A(G)$ ~10.5 ± 0.5, which indicates a doping of ≪100 meV.[57,58] After PVA deposition, FWHM(G) ~ 11.6 ± 1.2 cm$^{-1}$, Pos(2D) ~2676.0 ± 1.2 cm$^{-1}$, and $I(2D)/I(G)$ and $A(2D)/A(G)$ are ~4.3 ± 0.4 and ~9.0 ± 0.6, respectively. An almost identical range of FWHM(2D) is measured before and after PVA deposition (26.5 ± 1 cm$^{-1}$ and 26.4 ± 0.8 cm$^{-1}$). This indicates that PVA has no significant effect on the overall SLG quality, except for increased doping (~100 meV).[57,58] SLG on top of PVA presents larger values and spread of Pos(G) ~1585.2 ± 1.3 cm$^{-1}$ and lower FWHM(G) ~9.8 ± 1.4 cm$^{-1}$, indicating a higher level and variation of doping in the top SLG.

The rate of change of Pos(2D) and Pos(G) with strain is ruled by the Grüneisen parameters,[59] which relate the relative change in the peak positions in response to strain. Biaxial strain can be differentiated from uniaxial by the absence of G and 2D peak splitting with increasing strain.[54] However, at low (≲0.5%) strain, the splitting cannot be resolved.[59,60] Figure 4e plots the correlation between Pos(2D) and Pos(G). In general, both strain and doping influence Pos(G) and Pos(2D),[57−60] while strain does not influence $A(2D)/A(G)$. FWHM(G) can change due to strain inhomogeneities giving a distribution of slightly different Pos(G) in the area probed by the laser spot size or for a uniform strain, at the onset of the splitting of the G peak.[59] We use this approach to analyze the data in Figure 4. We first derive the doping from $A(2D)/A(G)$. We then consider Pos(G) as a function of Pos(2D). In undoped samples, Pos(G) and Pos(2D) are linked by the Gruneisen parameters.[59] Any deviation from the relation between Pos(G) and Pos(2D) expected for undoped samples can thus be assigned to the presence of both strain and doping. Finally, for doping <1/2 the energy of the G phonon (~100 meV),[61] the variation of Pos(G) is mostly due to strain,[57] which can then be derived. This approach is applied to analyze the fits in Figure 4 as follows. At linear fit to data, solid lines in Figure 4e, gives a slope $\Delta$Pos(2D)/$\Delta$Pos(G) ~ 1.44, ~1.52, ~1.65 for SLG on BPTEOS, SLG with PVA, and SLG on PVA, respectively. These values indicate a variation of both doping and strain within the mapped area, comparable to that of polycrystalline CVD-SLG encapsulated with hBN.[48] The presence (or coexistence) of biaxial strain cannot be ruled out. For uniaxial(biaxial) strain, Pos(G) shifts by $\Delta$Pos(G)/$\Delta\varepsilon$ ~ 23(60) cm$^{-1}$/%.[59,60,62] For intrinsic SLG ($E_F$ < 100 meV), the unstrained, undoped Pos(G) ~1581.5 cm$^{-1}$.[53,61] Our graphene on BPTEOS, SLG with PVA, and SLG on PVA has a mean Pos(G) ~1582.5, 1583.7, 1585.2 cm$^{-1}$. These values suggest a mean uniaxial (biaxial) strain $\varepsilon$ ~ 0.04% (~0.02%), ~ 0.1% (~0.04%), and ~0.13% (~0.05%), respectively.

We perform static and radio frequency (RF) characterizations, as detailed in Methods, in order to estimate the maximum $R_V$ and the BW. We bias the devices at 10 mV and measure the current in a two-terminal configuration as a function of the voltages applied to the two split gates ($V_{G1}$ and $V_{G2}$). The resistance map in Figure 5a shows four regions, defined by the different doping in the two sides of the junction. The quadrants correspond to finite $n$ in the gated regions, with the four possible doping configurations n–n, n–p, p–n, and p–p. The peak $R$ lines track the CNPs at 1.2 and 0.8 V for split gates 1 and 2, respectively. Considering the thickness of the PVA (120 nm) and the 14.5 dielectric permittivity (see Methods), at $V_{G1,G2}$ = 0 V, we estimate $n$ ~ 1 × 10$^{12}$ cm$^{-2}$ and ~7 × 10$^{11}$ cm$^{-2}$, for the two gated regions, in agreement with the Raman estimates.

Figure 5b is a $V_{Ph}$ map of the GPD, over the same $V_{G1}$–$V_{G2}$ range for an unbiased device coupled to a 1550 nm laser. While the resistance map has four gate voltage regions, the photoresponse shows a 6-fold pattern of alternating voltage as a function of $V_{G1}$ and $V_{G2}$, as emphasized by the dashed lines. The nonmonotonic variation of photoresponse as a function of gate voltage is reminiscent of the S dependence on $n$ in Figure 3c. The PV effect, due to the variation of gate potential along the channel,[32,63] is a photoconversion mechanism competing with PTE.[32,63] In general, and depending on device geometry and wavelength, both mechanisms can be present.[63] However, PV is generally weak in graphene p–n homojunctions with respect to PTE at zero bias,[32] even more so for samples having a narrow charge neutrality region (<100 meV) and high mobility (>10 000 cm$^2$ V$^{-1}$ s$^{-1}$).[32] In our case, the dominance of PTE is experimentally verified from the photovoltage/photocurrent maps in Figure 5b–d, showing six different zones of alternating photoresponse sign, as marked by the dashed lines and labels. This is the signature of the PTE effect, because it is due to variations of the $S$ difference between the two sides of the junction,[32] and cannot be attributed to PV, which would exhibit only two zones of different signs in the photoresponse map.[32]

The experimental Figure 5b is consistent with the simulated photovoltage map in Figure 3f, confirming that the SLG-PVA structure supports sharp variations of $S$ across the gated regions.

At a 700 $\mu$W input power, the maximum $V_{Ph}$ ~ 4 mV is found for the p–n and n–p regions in proximity of the neutrality lines.

Considering the various sources of loss, as detailed in Methods, the optical incident power is ~700 $\mu$W, yielding $R_V$ ~ 6 V W$^{-1}$. Figure 5c,d reports the photocurrent and responsivity maps for our GPDs, displaying the 6-fold pattern signature of PTE effect.[30,32] The doping can be tuned to be p-type or n-type, depending on the applied gate voltages, and hence p–n indicates the region of p-doping induced by the gate 1 ($V_{G1}$) and n-doping induced by gate 2 ($V_{G2}$), and so on, with n′ (p′) indicating stronger n-type (p-type) doping with respect to n (p).

Figure 6 plots the frequency response of the photodetector to an amplitude-modulated optical signal (see Methods). The response is flat up to 67 GHz, limited by our measurement setup, without evidence of signal roll-off, defined as the slope of the frequency response in the transition region between the passband (cutoff frequency of the flat electro-optical response) and stopband (cutoff frequency at which the electro-optical response is null).[64] Some ripples are present in the response, originating from cavity effects along the measurement cables, due to the presence of bias-tee and RF adapters in the setup, see Methods. The −3 dB level is defined by averaging the frequency response in the range 2−15 GHz, and using this as reference. Three separate frequency response measurements are performed. They show a similarly flat response up to 67 GHz, with no evidence of signal roll-off. The maximum theoretical PTE photodetector speed is related to the hot electron cooling time $t$ ~ 2−4 ps (ref 12), corresponding to 250−500 GHz.[24] Table 1 compares our results with existing reports.[14−16,23−25] This shows that our GPDs have the highest frequency to date, for zero bias devices. Our scalable technology is a key enabler for next-generation graphene-based transceivers.

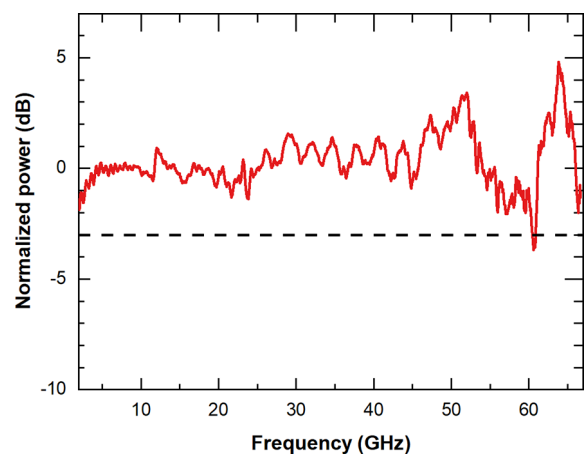

Figure 6. Frequency response of our GPD measured up to 67 GHz. The dashed line indicates the 3 dB drop of the electro-optical response. In the frequency range of the measurement, the roll off of the electro-optical response does not occur.

## CONCLUSION

We demonstrated ultrafast scalable double-SLG photodetectors integrated on a Si$_3$N$_4$ WG, using SLG as an active channel and split gates, separated by a PVA dielectric. Our devices operate with zero dark current, showing a flat frequency response up to 67 GHz without roll-off. The GPDs are fabricated using deterministically grown single-crystal SLG arrays. These and the fabrication process are compatible with back-end-of-line wafer-scale integration. The use of a polymer dielectric allows us to preserve the SLG quality, unlike conventional dielectrics, such as $Al_2O_3$ or $HfO_2$, whose deposition process typically degrades SLG.[36] This is also CMOS compatible, since the deposition of PVA poses a low risk of contamination and does not require a high temperature. Thanks to this dielectric we get carrier mobility >16,000 cm$^2$ V$^{-1}$ s$^{-1}$ in ambient conditions and without hBN encapsulation. The high carrier mobility is a key enabler for high-performing graphene-based photonics.[10] Thanks to the polymeric dielectric, the GPD channel has a narrow charge neutrality region (~35 meV) and high Seebeck coefficient, allowing our devices to operate in a PTE regime. Our GPDs do not require either a doped Si structure for gating[25] or nanoscale plasmonic structures,[15,24,25] making them ideal for large-scale integration on any photonic platform, thus a key building block for optical communications.

Table 1. Comparison of WG-Integrated GPDs

| ref | dominant mechanism | dark current (applied bias) | type of graphene | responsivity | bandwidth |
|---|---|---|---|---|---|
| this work | PTE | none | CVD | $R_V$ = 6 V W$^{-1}$ | >67 GHz (setup-limited) |
| 14 | PV | none | CVD | $R_I$ = 0.016 A W$^{-1}$ | 41 GHz |
| 15 | PTE | none | flake | $R_V$ = 4.7 V W$^{-1}$ | 18 GHz |
|  | PV/PBM at 0.4 V bias | 0.4 V |  | $R_I$ = 0.17 A W$^{-1}$ | (pulse measurement) |
| 16 | PBM | 1 V | CVD | $R_I$ = 0.001 A W$^{-1}$ | 76 GHz |
|  |  |  |  | $R_V$ = 0.13 V W$^{-1}$ |  |
| 17 | n/a | 1 V | CVD | $R_I$ = 0.18 A W$^{-1}$ at 0.5 V bias | >128 GHz at 1 V bias (setup-limited) |
|  |  | 0.5 V |  | $R_V$ = V W$^{-1}$ at 0.5 V bias |  |
| 18 | PBM | 4 mA | CVD | $R_I$ = 0.5 A W$^{-1}$ | 110 GHz |
|  |  | (0.4 V) |  |  |  |
| 19 | PV | 1.6 V | CVD | 0.36 A W$^{-1}$ | 110 GHz |
| 23 | PTE | none | flake | $R_I$ = 0.078 A W$^{-1}$ | 42 GHz |
|  |  |  |  | ($R_I$ = 0.36 A W$^{-1}$ at 1.2 V bias) |  |
| 24 | PTE | none | CVD | $R_V$ = 12 V W$^{-1}$ | 42 GHz |
| 25 | PTE | none | flake | $R_V$ = 3.5 V W$^{-1}$ | 65 GHz |
|  |  |  |  | $R_I$ = 0.035 A W$^{-1}$ |  |

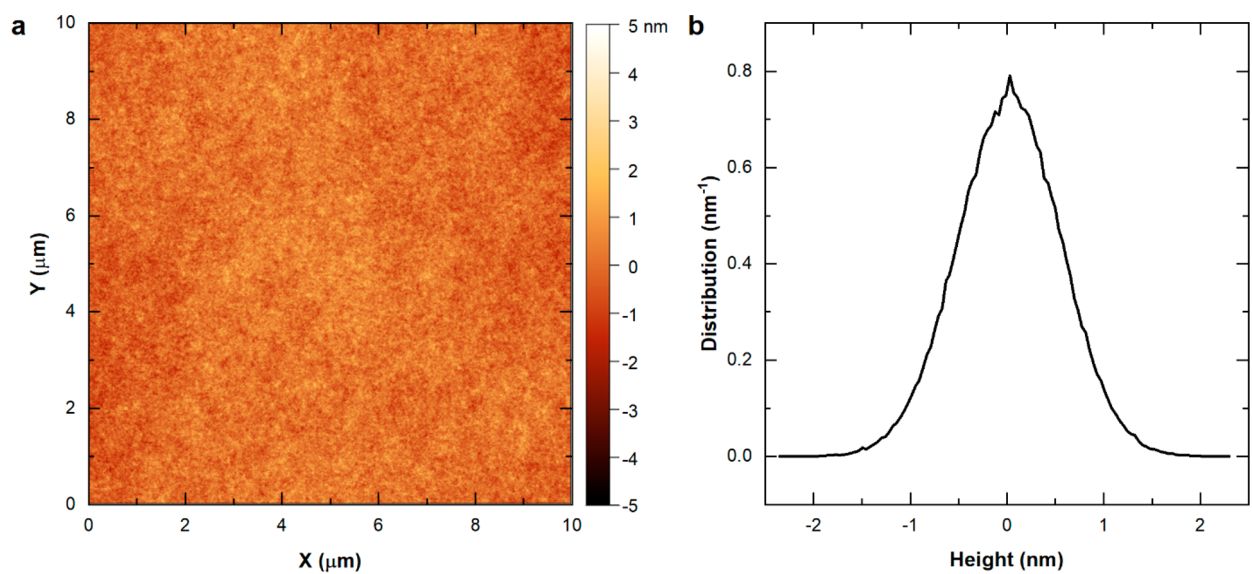

Figure 7. (a) AFM image of 10 × 10 $\mu$m$^2$ surface of PVA. (b) Height distribution of the area in panel (a).

## METHODS

**Fabrication.** The devices are fabricated by depositing 260 nm $Si_3N_4$ on 15 $\mu$m $SiO_2$ via low-pressure CVD. The 1.5 $\mu$m-wide WGs are defined in $Si_3N_4$ using EBL and reactive ion etching. The surface is then planarized with BPTEOS with a final thickness of 25 nm.

Our semidry transfer procedure uses a micrometric stage to laminate the SLG crystals onto the target substrate in a controlled manner, thus avoiding the formation of wrinkles, as shown in the high-contrast optical micrograph in Figure 2a. The SLG single crystals are grown deterministically in arrays,[42] Figure 2b. For a ~1.5 × 1.5 cm$^2$ substrate, a well-ordered matrix of ~5000 crystals is grown. Depending on the geometry of the target photonic chip, ~200−300 of these crystals can be aligned and transferred to WGs, as shown in Figure 2c. In a typical fabrication run, ~20 crystals are processed into GPDs, which demonstrates the scalability of this approach. Reference 42 showed by selected area electron diffraction that the crystals have a single orientation. The Raman spectrum of a representative crystal in Figure 4a shows no D peak, indicating negligible defects.[54,55] Our single crystals encapsulated in hBN have a room-temperature mobility ~1.3 × 10$^5$ cm$^2$ V$^{-1}$ s$^{-1}$ (at a charge density ~10$^{11}$ cm$^{-2}$) and a low-temperature (4.2 K) mobility >6 × 10$^5$ cm$^2$ V$^{-1}$ s$^{-1}$ (at a charge density ~10$^{11}$ cm$^{-2}$) and show signatures of electronic correlation, including the fractional quantum Hall effect, as discussed in ref 65.

Typical SLG/PVA/SLG FETs have gate breakdown voltages up to 0.4 MV cm$^{-1}$. However, overlapping the top split gate with the bottom metal contacts can give values ~1 order of magnitude lower. Therefore, we place the split gate structure between the metal contacts. The design requires at least 1 $\mu$m of lateral separation between gates and metal source−drain contacts, where the spin-coated dielectric is more planar. Thus, the minimum source−drain distance used in our detectors is 10 $\mu$m.

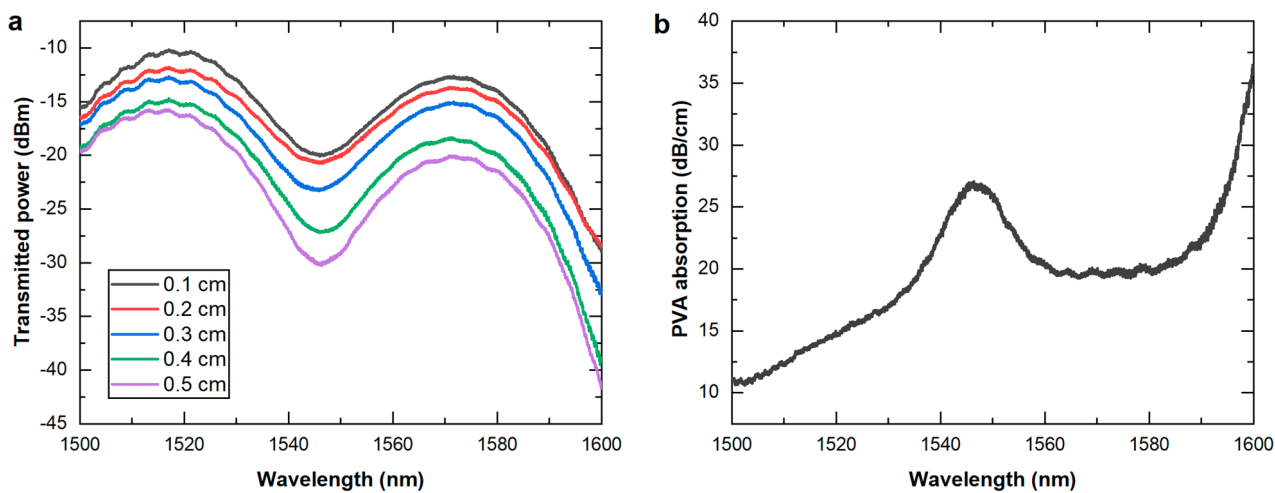

Figure 8. (a) Absorption of PVA coating at various lengths on a $Si_3N_4$ WG as a function of wavelength. (b) Calculated loss of PVA coating per unit length as a function of wavelength.

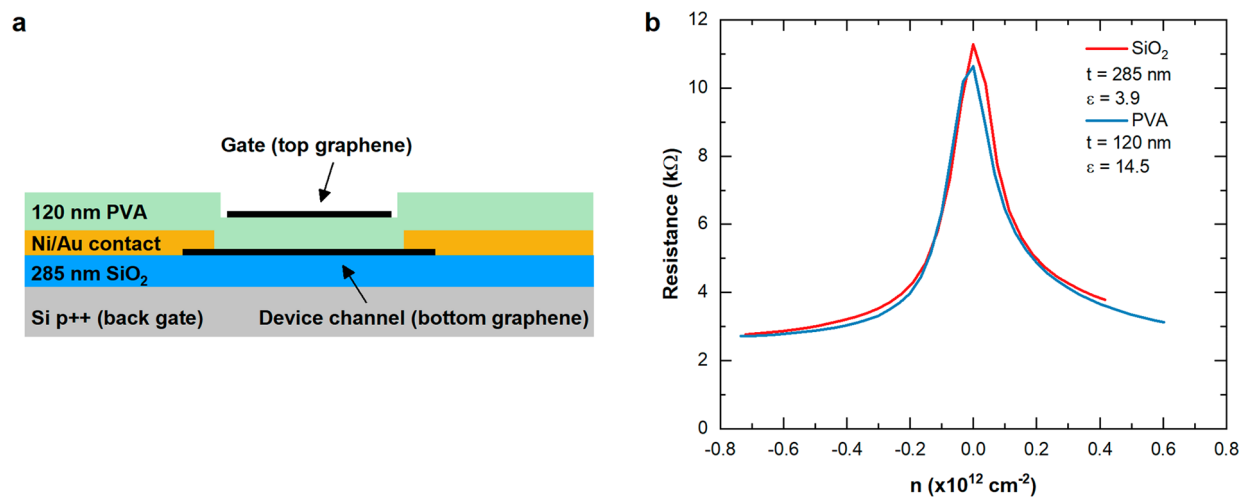

Figure 9. (a) A cross-section of a double-gate test structure. (b) Comparison of field-effect characteristics when using bottom gate ($SiO_2$ dielectric) and top gate (PVA dielectric).

**PVA Preparation and Characterization.** The PVA solution is prepared by dissolving PVA powder (Sigma-Aldrich, 23−75 kDa, 89% hydrolyzed) in DI water (∼ 18.2 M$\Omega$ cm) and then passing the solution through a 0.22 $\mu$m filter.

Atomic force microscopy (AFM) (Bruker Dimension Icon) in tapping mode is used to estimate the PVA roughness prior to transfer of the top SLG. A representative AFM image is in Figure 7a. Figure 7b plots the height distribution of 10 × 10 $\mu$m$^2$ of spin-coated PVA, indicating an RMS roughness ∼0.53 nm.

Cut-back measurements[66] are performed to estimate the optical absorption of PVA spin-coated on WGs. We use five WGs with different lengths ranging from 1 to 5 mm, on top of which 120 nm-thick PVA is deposited by spin-coating. The transmission of the WGs is measured in the 1500−1600 nm range in Figure 8a. At 1550 nm, the extracted losses are ∼2.5 dB mm$^{-1}$, Figure 8b.

Refs 67−69 reported relative permittivity $\varepsilon_r$ for thin PVA films (thickness <1 $\mu$m), ranging from 5[67] up to well above 20.[69] To determine the $\varepsilon_r$(PVA) for our devices, we fabricate double-gated FETs on highly p-doped Si wafers (<0.005 $\Omega$ cm) with 285 nm SiO$_2$. SLG growth, transfer, and fabrication are identical to those used for the GPDs in Figure 1. A profile of such structure is shown in Figure 9a.

Field-effect measurements are performed on a double-gated device by sweeping top (bottom) gate and keeping the bottom (top) gate fixed. From the applied gate voltage we get n = $\varepsilon_0 \varepsilon_r V_g / (t \times q)$,[70] where $\varepsilon_0$ = 8.85 × 10$^{-12}$ F m$^{-1}$ is the permittivity of free space, $\varepsilon_r$ is the relative permittivity, $t$ is gate dielectric thickness (285 nm for SiO$_2$ and 120 nm for PVA), and $q$ = 1.60 × 10$^{-19}$ C is the elementary charge. Using $\varepsilon_r$(SiO$_2$) = 3.9 (ref 71), $\varepsilon_r$(PVA) is adjusted to find a good agreement of transfer curves obtained using the two gates, giving ∼14.5 (Figure 9b).

**DC Characterization.** DC measurements are performed by contacting the source−drain channel and the two split gates using micromanipulators with DC needle probes to Keithley 2602 source/measure units. A resistance map of each device is obtained by applying 10 mV to the source−drain channel, and sweeping the voltages applied to the split gates while measuring the current flow. The static photoresponse is measured by coupling a continuous wave (CW) laser at 1550 nm with a single-mode optical fiber and a grating coupler. A polarization controller is used to match the polarization of the optical field at the grating coupler. Photovoltage maps are obtained by imposing zero current between source and drain electrodes and sweeping the voltages applied to both SLG gates.

Due to the SLG band structure and zero band, SLG p−n homojunctions do not work like conventional diodes.[26] Due to the interband tunneling in proximity of the junction,[72,73] whenever a bias (0.1−1 V) is applied between source and drain contacts, a large (∼mA) current flows regardless of applied bias sign.[73] Therefore, a reverse bias region, where the dark current can be suppressed, does not exist. The only possible condition to avoid a significant (∼mA) dark current is zero bias operation. For this reason, the measurement of the photocurrent at nonzero bias is not done. Furthermore, the goal is to minimize power consumption, hence, there is no incentive in operating the devices at high voltage.

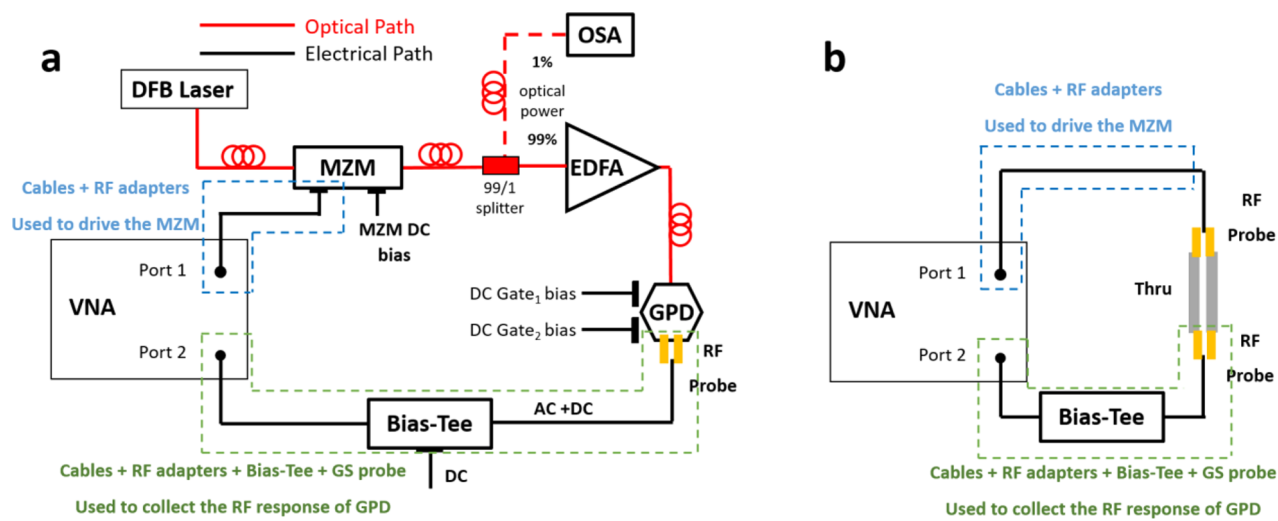

Figure 10. Block diagram of the instrumental setup for (a) RF characterization of the GPD and (b) calibration of the RF measurement.

The output of the laser source is amplified by an erbium doped fiber amplifier (EDFA), giving a final power ∼17 dBm (50 mW). The optical power at the output of the polarization controller is ∼15 dBm. The optical power measured from the output grating coupler is ∼ −25 dBm. According to simulations of the optical mode of the WG with the detector stack, the SLG stack is expected to absorb ∼0.09 dB $\mu m^{-1}$, contributing 7.2 dB to the total loss (for a device length ∼80 $\mu$m). The remaining 32.8 dB losses are attributed to grating couplers and propagation loss. Grating couplers introduce a 7 dB loss each. The 18.8 dB propagation loss is due to the ∼25 nm BPTEOS cladding on the $Si_3N_4$ WG. Any residues due to the various fabrication steps can introduce propagation loss. By assuming homogeneously distributed losses along the WG, we estimate a ∼16.4 dB loss between polarization controller and GPD. The optical power reaching the GPD is therefore ∼ −1.4 dBm (700 $\mu$W). Negligible propagation losses from the grating coupler to the device could be achieved by using thick (>1 $\mu$m) cladding, which can lead to propagation losses as low as 1.5 dB cm$^{-1}$ for $Si_3N_4$ WGs.[74] Selective removal of oxide could be used to reduce the cladding to ∼20 nm in the device areas, which is required for the evanescent coupling between GPD and WG mode, as shown by the simulations in Figure 3d.

**RF Characterization.** To measure the optoelectronic bandwidth of the GPD up to 67 GHz, an electrical Vector Network Analyzer (VNA) is employed. We use the measurement scheme of Figure 10a. A distributed feedback (DFB) laser generates a CW optical signal at 1.55 $\mu$m. The laser optical power is fed into a Mach−Zehnder modulator (MZM) with 40 GHz BW. The MZM is used to amplitude modulate the DFB CW laser from 2 to 67 GHz, with a sinusoidal signal. The electrical modulation to the MZM is delivered by one of the ports of an electrical VNA (port 1 in Figure 10a). The modulated laser beam is then amplified by an EDFA (Keopsys, with a tunable optical output from 22 to 30 dBm) and guided with an optical fiber up to the GPD, where it is coupled by a grating. The GPD is electrically connected to a second port of the VNA (port 2 in Figure 10a) through electrical probes for on-wafer measurements (Cascade Microtech APC67). Consequently, the modulating electrical frequency at port 1 of the VNA is swept, which is delivered to the MZM, up to 67 GHz. For each frequency sweep step, the GPD photoresponse is measured at port 2. During characterization, the frequency response of the MZM up to 67 GHz is measured. This allows us to correct the experimental results for various frequency-dependent losses introduced by the MZM. To do so, we use an Optical Signal Analyzer (OSA) (Yokogawa AQ6370D) to monitor the modulation depth (red dashed line in Figure 10) at each frequency sweep step of port 1. The optical power at the output of the MZM is $P(t) = P_0(1 + m \sin 2\pi f_0 t)$, where $P_0$ is the constant part of the modulated optical power, $f_0$ is the RF frequency applied to the MZM, and $m$ is the modulation depth.[75] As shown in Figure 10b, we perform a short-open-load-through calibration[76] to eliminate the contribution of the cables to the frequency response. Due to the presence of ripples also in the low-frequency region, we average the frequency response in the range of 2−15 GHz and use this as a reference level to define the −3 dB.

**Simulations.** We calculate the responsivity by adapting the model from ref 32 to the case of a WG-integrated GPD:

$$\nabla^2 T_e + \frac{1}{L_C^2}(T_e - T_0) = \frac{P_{absorbed}(x, z)}{K_e h_g} \quad (2)$$

$$\vec{J} = -\sigma \nabla V + \sigma S[\mu(x)] \nabla T_e, \quad \nabla \cdot \vec{J} = 0 \quad (3)$$

where $L_C$ is the cooling length[30,32] and accounts for the hot electron relaxation,[30,32] $P_{absorbed}(x,z)$ [W m$^{-2}$] is the density of absorbed optical power in SLG, where the photocurrent is generated, $K_e$ is the electron thermal conductivity, related to the SLG channel electrical conductivity through the Wiedemann−Franz law, $h_g$ = 0.34 nm is the thickness of an ideal SLG,[77] $\vec{J}$ is the current density [A m$^{-1}$], and $\mu(x)$ is the chemical potential at position $x$. $P_{absorbed}$ is the spatial profile of the absorbed optical power density in SLG, extracted from a simulation of the fundamental quasi-TE mode profile of the $Si_3N_4$ WG (1200 nm × 260 nm, $\lambda_0$ = 1550 nm) with the detector stack (see Figure 3d). Eq 3 is solved using a commercial finite elements method solver.

In order to extract $L_C$, we realize a non-optimized GPD. We compare the measured $R_V$ with those simulated for devices having the same geometry. Simulation parameters like $S$ and channel resistance are obtained by the measurements of FET resistance curves of SLG with PVA as gate dielectric. $L_C$ is varied in a range 0.1−1 $\mu$m, since the values reported in literature for this parameter vary from 140 nm[78] to 1 $\mu$m.[23] The value extracted, 130 nm, is compatible with that in polycrystalline CVD SLG encapsulated in hBN.[78]

**Seebeck Coefficient.** In order to calculate $S$, we use the field-effect measurement in Figure 11a to extract $n^*$. The model in ref 46 is used to fit the contact resistance, $R_C$, the field-effect mobility $\mu_{FE}$ and $n^*$ (Figure 11b). We also use the model in refs 44 and 45. We find good agreement of the parameters extracted by the two methods.

From ref 46, $R$ can be written as

$$R = R_C + \frac{N_{sq}}{q\mu_{FE}\sqrt{n^{*2} + n[V_{top\,gate}]^2}} \quad (4)$$

and

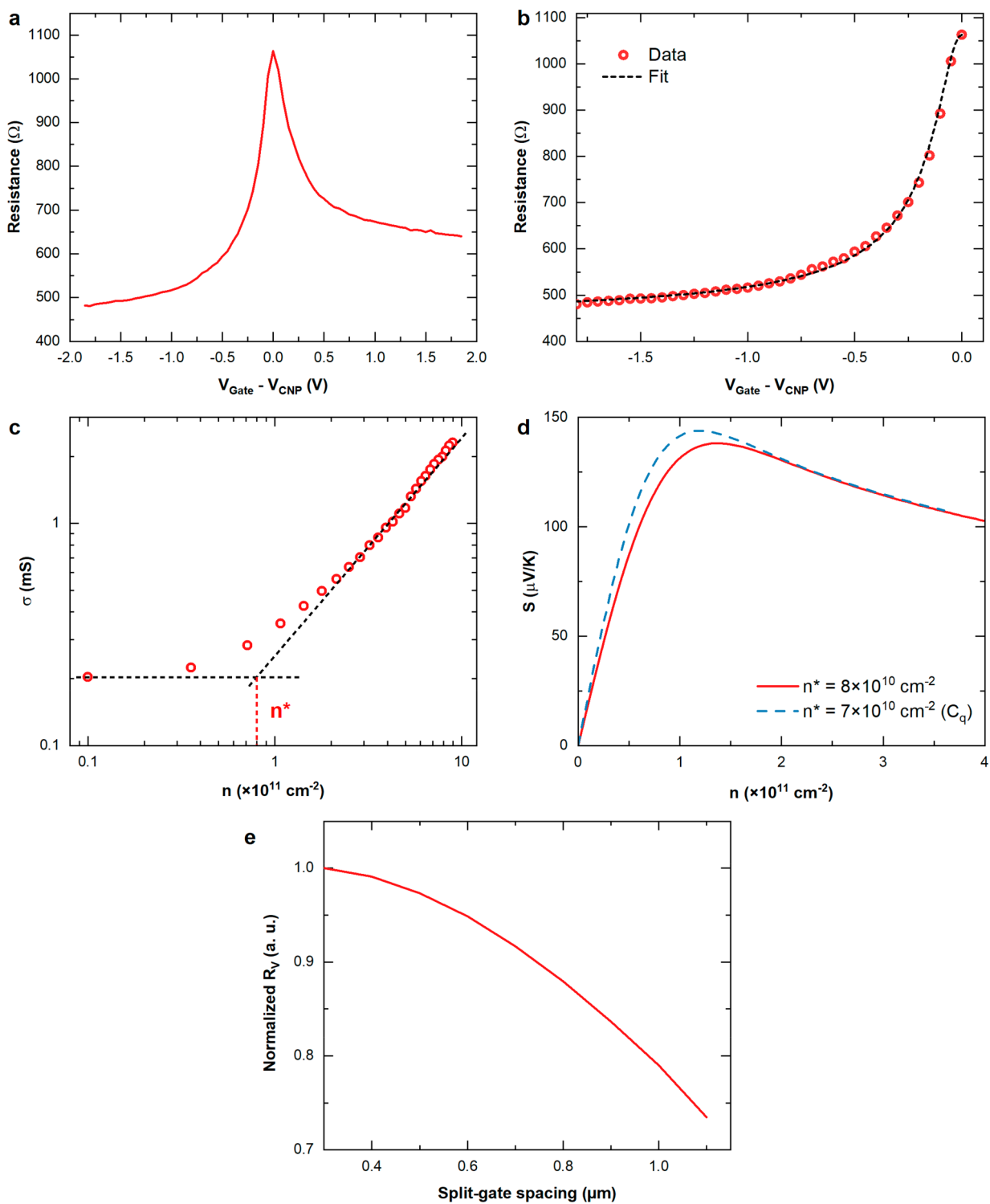

Figure 11. (a) Field-effect response of SLG with PVA gate dielectric. (b) Fit with eq 4. (c) Fit with eq 7. (d) S as a function of n for n* ∼ 8 × 10¹⁰ cm⁻² (red curve) and n* ∼ 7 × 10¹⁰ cm⁻²(blue dashed curve). (e) Simulation of maximum responsivity as a function of split gate spacing, normalized to $R_V$ (200 nm).

$$\sigma = q\mu_{FE}\sqrt{n^{*2} + n[V_{gate}]^2} = q\mu_{FE}\sqrt{n^{*2} + C_{gate}^2 V_{gate}'^2/q^2} \quad (5)$$

where $N_{sq} = L/W$ is the SLG channel aspect ratio, $q$ is the elementary charge, $\sigma$ is the conductivity, $V_{CNP}$ is the voltage at CNP, $C_{gate}$ is the gate capacitance, and $V_{gate}' = V_{gate} - V_{CNP}$.

From eq 5, for $n \gg n^*$, the conductivity can be approximated as

$$\sigma \sim q\mu n \quad (6)$$

$$\log(\sigma) \sim \log(n) + \log(q\mu) \quad (7)$$

Therefore, $\mu_{FE}$ can be extracted by the linear fit in Figure 11c. We get $n^* \sim 8.2 \times 10^{10}$ cm⁻² and $\mu_{FE} \sim 16,000$ cm² V⁻¹ s⁻¹. The model of ref 33 is then used to calculate $S$, as in Figure 11d. We consider screened charged impurities as the main scattering mechanism at low carrier density.[33,79−81] Because of an inhomogeneous dielectric environment, due to the different permittivities of substrate and superstrate, an effective dielectric constant $\varepsilon_r = (\varepsilon_{PVA} + \varepsilon_{SiO_2})/2$[82] is used. We assume the impurity plane at 0 nm from the SLG. The effective medium theory[33,83] is used to take into account random potential fluctuations in proximity of the Dirac point that break up the density landscape in electron−hole puddles.[80,84] We also estimate the value of $n^*$ by taking into account the quantum capacitance $C_q$. We get $n[V_{gate}]$ from ref 46:

$$V_{gate} - V_{CNP} = \frac{q}{C_{gate}}n + 2\frac{\hbar v_F}{q}\sqrt{\pi n} = qn\left(\frac{1}{C_{gate}} + \frac{1}{C_q}\right) \quad (8)$$

where $C_q = 2q^2\sqrt{n}/(\hbar v_F\sqrt{\pi})$,[85] $\hbar = 1.05 \times 10^{-34}$ J s is the reduced Planck constant, and $v_F = 1 \times 10^6$ m s$^{-1}$ is the Fermi velocity. The obtained value $n^* \sim 7 \times 10^{10}$ cm$^{-2}$ does not lead to a significant change to the calculated $S$, as shown in Figure 11d (blue dashed curve).

In order to study various split-gate geometries, the photoresponse is simulated by varying the split gate spacing between 0.2 and 1.1 $\mu$m. We assume a step $S$ spatial profile (as in Figure 3e) and vary the width of the split gate separation where $S = 0$. Though such an abrupt transition of chemical potential and $S$ is not expected between the different zones, this allows us to get a lower bound estimation of $R_V$, as the step profile represents the worst case for evaluation of $R_V$ as a function of gap width. At 1 $\mu$m separation, the $R_V$ is ∼78% of $R_V$ obtained with a 200 nm gap, as shown in Figure 11e. This is due to the ∼1.5 $\mu$m spatial extension of the optical mode sustained by the Si$_3$N$_4$ WG. The resulting $T_e$ profile has its maximum ∼500 nm from the center of the channel (see Figure 3e).

## AUTHOR INFORMATION

**Corresponding Authors**

**Camilla Coletti** − *Center for Nanotechnology Innovation @ NEST and Graphene Labs, Istituto Italiano di Tecnologia, 56127 Pisa, Italy;* orcid.org/0000-0002-8134-7633; Email: camilla.coletti@iit.it

**Marco Romagnoli** − *Photonic Networks and Technologies Lab, CNIT, 56124 Pisa, Italy;* Email: marco.romagnoli@cnit.it

**Authors**

**Vaidotas Mišeikis** − *Center for Nanotechnology Innovation @ NEST and Graphene Labs, Istituto Italiano di Tecnologia, 56127 Pisa, Italy; Photonic Networks and Technologies Lab, CNIT, 56124 Pisa, Italy;* orcid.org/0000-0001-6263-4250

**Simone Marconi** − *Photonic Networks and Technologies Lab, CNIT, 56124 Pisa, Italy; TeCIP Institute, Scuola Superiore Sant'Anna, 56124 Pisa, Italy*

**Marco A. Giambra** − *Photonic Networks and Technologies Lab, CNIT, 56124 Pisa, Italy; TeCIP Institute, Scuola Superiore Sant'Anna, 56124 Pisa, Italy;* orcid.org/0000-0002-1566-2395

**Alberto Montanaro** − *Photonic Networks and Technologies Lab, CNIT, 56124 Pisa, Italy*

**Leonardo Martini** − *Center for Nanotechnology Innovation @ NEST, Istituto Italiano di Tecnologia, 56127 Pisa, Italy*

**Filippo Fabbri** − *Center for Nanotechnology Innovation @NEST and Graphene Labs, Istituto Italiano di Tecnologia, 56127 Pisa, Italy;* orcid.org/0000-0003-1142-0441

**Sergio Pezzini** − *Center for Nanotechnology Innovation @NEST and Graphene Labs, Istituto Italiano di Tecnologia, 56127 Pisa, Italy;* orcid.org/0000-0003-4289-907X

**Giulia Piccinini** − *Center for Nanotechnology Innovation @ NEST and Graphene Labs, Istituto Italiano di Tecnologia, 56127 Pisa, Italy; NEST, Scuola Normale Superiore, 56126 Pisa, Italy*

**Stiven Forti** − *Center for Nanotechnology Innovation @NEST, Istituto Italiano di Tecnologia, 56127 Pisa, Italy*

**Bernat Terrés** − *ICFO - Institut de Ciencies Fotoniques, the Barcelona Institute of Science and Technology, 08860 Castelldefels, Spain*

**Ilya Goykhman** − *Technion - Israel Institute of Technology, Technion City 3200003, Haifa, Israel;* orcid.org/0000-0002-8833-9193

**Louiza Hamidouche** − *Thales Research and Technology, 91767 Palaiseau, France*

**Pierre Legagneux** − *Thales Research and Technology, 91767 Palaiseau, France*

**Vito Sorianello** − *Photonic Networks and Technologies Lab, CNIT, 56124 Pisa, Italy*

**Andrea C. Ferrari** − *Cambridge Graphene Centre, Cambridge University, Cambridge CB3 0FA, United Kingdom;* orcid.org/0000-0003-0907-9993

**Frank H. L. Koppens** − *ICFO - Institut de Ciencies Fotoniques, the Barcelona Institute of Science and Technology, 08860 Castelldefels, Spain; ICREA, Institució Catalana de Recerça i Estudis Avancats, Barcelona 08010, Spain;* orcid.org/0000-0001-9764-6120

Complete contact information is available at:
https://pubs.acs.org/10.1021/acsnano.0c02738

**Author Contributions**

◇These authors contributed equally.

**Notes**

The authors declare no competing financial interest.

## ACKNOWLEDGMENTS

We acknowledge funding from the European Union Graphene Flagship under grant agreement nos. 785219 and 881603, ERC grants Core2 and Core3, GSYNCOR, and EPSRC grants EP/L016087/1, EP/K01711X/1, EP/K017144/1, EP/N010345/1. F.H.L.K. acknowledges financial support from the Government of Catalonia trough the SGR grant, and from the Spanish Ministry of Economy and Competitiveness, through the "Severo Ochoa" Programme for Centres of Excellence in R&D (SEV-2015-0522), support by Fundacio Cellex Barcelona, Generalitat de Catalunya through the CERCA program, and the Mineco grants Plan Nacional (FIS2016-81044-P) and the Agency for Management of University and Research Grants (AGAUR) 2017 SGR 1656.

## NOTE ADDED AFTER ASAP PUBLICATION

Due to production error, figures were misprocessed in the version published on August 21, 2020 and were correctly restored on August 24, 2020.